\documentclass[12pt]{revtex4}
\usepackage[dvips]{graphicx,color}
\def\bea{\begin{eqnarray}}
\def\eea{\end{eqnarray}}
\def\be{\begin{equation}}
\def\ee{\end{equation}}
\def\nn{\nonumber}

\def\G{\Gamma}

\def\l{\lambda}
\def\L{\Lambda}
\def\m{\mu}
\def\n{\nu}

\def\r{\rho}
\def\s{\sigma}

\def\t{\tau}

\def\d{\delta}
\def\ve{\varepsilon}

\begin{document}

\title{The gravitational path integral and the trace of the diffeomorphisms}

\vspace{1cm}

\author{\sc Arundhati Dasgupta}
\affiliation{
  Department of Physics and Astronomy, University of Lethbridge, Lethbridge, Canada T1K 3M4}
\email{arundhati.dasgupta@uleth.ca}
\begin{abstract} 
I give a resolution of the conformal mode divergence in the Euclidean gravitational path-integral by
isolating the trace of the diffeomorphisms and its contribution to the Faddeev-Popov measure.
\end{abstract}

\maketitle

\section{Introduction}
The `path-integral' in quantum gravity is defined as a functional `integral' over metrics.
 However, this `integral' has not been completely obtained due
to technical difficulties including the non-Gaussian nature of the integral as a function
of the metric and the unboundedness of the Euclidean action \cite{hawk} which appears in the `integrand'.
There has been considerable research in-order to solve some of the problems, and in this paper
we discuss the problem of the unboundedness of the Euclidean action. The Euclidean path integral is defined thus
$$\int {\cal D} g_{\m \n} \exp (-\Gamma_{\rm classical}). $$
Where $g_{\m \n}$ represents a positive definite metric and $\Gamma_{\rm classical}$ is Einstein's action for gravity
given by $\frac{-1}{16\pi G}\int\sqrt {\rm det \ g}~ R ~d^4x$. Here $\rm det \ g $ is the determinant of the
metric $g_{\m \n}$ , and $R$ is the associated scalar curvature. The action can be written in terms of the conformal mode or the scale factor of the metric and a set of conformal
equivalence class of metrics. The conformal mode $\phi$ of the metric can be isolated by writing $g_{\m \n}=e^{2\phi}\bar g_{\m \n}$
by using the Yamabe conjecture, where the conformal mode is fixed by requiring that the metric $\bar g_{\m \n}$ has a constant Ricci scalar. 
The action written in terms of these variables can be arbitrarily large and negative for a rapidly fluctuating conformal mode as the conformal mode gives a term $\frac{-3}{8\pi G}\int d^4 x \  e^{2 \phi} \sqrt{\bar g} (\bar\nabla\phi)^2$ to the Euclidean action ($\bar \nabla$ is a covariant derivative wrt the metric $\bar g_{\m \n}$). 
The term has the square of the derivative of the conformal mode, and thus the sign of the kinetic term is fixed for real $\phi$. Metric configurations 
with unbounded derivatives of the conformal mode will contribute to make the Einstein action arbitrarily negative.
Classically one can argue for positive action metrics, but in case of a quantum calculation, where the path-integral
includes integration over arbitrary non-classical metrics, configurations with large negative actions would exist. 
The Euclidean path-integral which has the exponential of the negative of the Euclidean action is thus potentially divergent.
 Previous attempts to examine this particular problem
\cite{schl,schl1,mm,adgloll}, have concluded that the perturbative gravitational path integral when written in terms of the `physical variables'
has a positive definite effective action.  
This `identification' of physical variables done by factoring out the diffeomorphism group from the measure is well defined in the perturbative regime or for metrics which are fluctuations over a classical metric. The identification of physical variables for the non-perturbative metrics remained a very difficult task. 

In a path-integral, the physical metrics are identified in the measure in the integral by factoring out redundant configurations related by the diffeomorphism group. 
A set of metrics is identified as the physical metrics fixed by certain gauge conditions, and any other metric related to the physical set by diffeomorphisms is factored out of the measure. The physical measure now includes a Jacobian
of the diffeomorphism transformations which relate the redundant metrics to the physical metrics. This Jacobian is known as the Faddeev-Popov determinant. The physical set of metrics
are known as `gauge fixed metrics' and the evaluation of the Faddeev-Popov determinant as a function of the metrics completes this procedure.

 In \cite{adgloll} we used a formalism given in \cite{mm} to show that the Faddeev-Popov determinant for physical metrics defined using proper-time gauge constrains the metrics
with large negative actions. The proper-time gauge was used as the path-integral could be defined using Lorentzian metrics and the analytically continued
Euclidean metrics in the path-integral would have causal histories. The result was verified in the perturbative regime using an explicit calculation \cite{adgloll}. The non-perturbative regime
was not calculated explicitly.  
In this paper, I define a new way of obtaining the physical measure and give an explicit non-perturbative calculation. 
This is achieved by writing the diffeomorphism transformations as comprising of a traceless part and a trace part. The conformal mode transforms due to the trace part
of the diffeomorphisms, and I concentrate on the trace of the diffeomorphisms by parametrising the trace of the diffeomorphism by a scalar field. The terms in the measure which are due to factoring out the diffeomorphism generated by the scalar field
are then isolated. Thus in some sense, I am using the Faddeev-Popov procedure only for the `scale' or conformal sector of the metric.
The Jacobian of the pure scale transformations is a scalar determinant which is then evaluated using heat kernel techniques for arbitrary non-perturbative metrics. This non-perturbative determinant 
makes the classical negative action positive. This result is true in any gauge and it appears that the resolution of the unboundedness
of the Euclidean action arises from factoring out just the trace of the diffeomorphisms.

To summarise:\\
1)The gravitational action can assume arbitrarily negative values due to the kinetic term of the
scale factor or the conformal mode of the metric. \\ 
2)The measure has redundant degrees of freedom due to the diffeomorphism group, and finding physical coordinates by factoring out the diffeomorphism group gives the Faddeev-Popov determinant. This comprises of scalar, vector and tensor determinants, and isolating the scalar determinant
achieves the resolution of the conformal mode problem discussed in this paper.\\
3)We use the formalism of \cite{bbm}, where the physical metric is chosen by imposing gauge conditions on the metric $\bar g_{\m \n}$ and the Yamabe condition ensures that orbits of $\phi$ are transversal to the
coordinates on the gauge slice. This approach is slightly different than the usual Faddeev-Popov ghost gauge fixing procedure were no such splitting of the metric is done.
 Thus the details of the gauge condition are encoded in the tensor Faddeev-Popov determinant which is due to gauge fixed $\bar g_{\m \n}$. The functional
form of the scalar determinant (which originates from factoring out the trace of the diffeomorphisms) is same in any gauge, and in this paper this is obtained
for arbitrary non-perturbative metrics using heat kernel techniques.
Thus for the purposes of this paper, the details of the gauge are not relevant for the resolution of the unboundedness of the Euclidean action.

The actual calculation is executed in the following way:
The path integral comprises of the measure ${\cal D} g_{\m \n}$ and the integrand $\exp(-\G_{\rm classical}) $ where $\G_{\rm classical}$ is the classical gravitational action.
We find the physical measure by dividing the given measure ${\cal D} g_{\m \n}$ by diffeomorphism group of the manifold ${\cal M}$, $Diff({\cal M})$. The physical measure has a Faddeev-Popov determinant, which is written in exponentiated form in the path-integral 
as $\exp (-\G_1)$. Thus
\be
\int \frac{{\cal D} g_{\m \n}}{Diff (\cal{M})} \ \exp(-\G_{\rm classical})= \int {\cal D} g_{\m \n}^{\rm phys} \exp(-\G_1-\G_{\rm classical}).
\ee
This is an integral over physical metrics with the weight corresponding to an effective action $\G_{\rm effective}= \G_1 + \G_{\rm classical}$.
The contribution from the physical measure $\G_1$ can be split into that due to the 
the trace of the diffeomorphisms $\G_{\rm trace}$, and the remaining ($\Gamma_{\rm traceless}$). 
 To find $\G_1$, 
 I explicitly find a heat kernel of a non-Laplacian operator, which appears in the scalar Faddeev-Popov determinant. 
In the heat kernel calculation of the determinant, 
 the regulator independent or `finite term' is extracted,
and is found to have the exact functional form as the classical action, but with a flipped sign in the regime where the curvature of space-time is slowly varying or $\nabla^{\mu}R_{\m \n}\sim 0$. In the regime
where space-time is strongly curved or $\nabla^{\mu}R_{\m \n}>>1$ the conformal mode term is rendered positive but there appear higher order curvature terms ($\bar R^2$ and $\bar \nabla^{\mu} \bar R_{\m \n}$) contributing to the effective action. Thus the positive $\G_1$ term when added to the negative $\sqrt{g} R$ term in the classical action renders the effective action positive definite. 
Thus this
is not a counter term prescription where a regulator dependent badly divergent term is controlled by a counter term to get the physical parameters finite. The measure in the path-integral includes
naturally a term which has the same functional form as the classical action with a opposite sign, and it reverses the overall sign of the $\int \sqrt{g} R $ term of the 
effective action. To clarify further this effective action is not obtained by integrating out matter fields coupled to gravity, but it is obtained by 
evaluating the determinants in the measure as a function of the physical metric.

In the next section the conformal mode problem in gravity is discussed, in section III the Faddeev-Popov measure is derived, in section IV, the resolution
of the conformal mode divergence is obtained for the perturbative case as well as the non-perturbative case. The non-perturbative case includes
the calculation of the heat kernel of the non-Laplacian operator, the details of which are derived in the Appendix.

\section{The Conformal mode in Gravity}
The Euclidean gravitational action or the action for positive definite metric comprises of 
\be
S= -\frac{1}{16\pi G} \int \sqrt{\det g}~ R  ~d^4x 
\ee
In \cite{hawk} Hawking showed that if there is a particular decomposition of the metric of the form;
\be
g_{\mu \nu} = e^{2\phi} \bar g_{\mu \nu}
\ee
where $e^{2\phi}$ is a conformal factor associated with the metric, and $\bar g_{\mu \nu}$ has a constant Ricci curvature $\bar R$, 
then the gravity action reduces to
\be
S= -\frac{1}{16 \pi G} \int  d^4 x  \ e^{2\phi} \sqrt{\bar g} \left[\bar R + 6 (\bar\nabla\phi)^2\right].
\ee
The kinetic term of the conformal mode is positive definite, and hence the Euclidean action can assume as
negative values as possible. This pathology can be assumed to be a signature of presence of redundant degrees of freedom, 
and indeed diffeomorphically related metrics in the measure are redundant. So, in this paper we investigate the path-integral written in terms of physical variables. 
To identify
physical metrics, one takes a `gauge-slice' in metric space, and other metrics related by pure diffeomorphism are factored out of the measure. This procedure, known
as gauge fixing, leads to an effective action, re-written only in terms of the physical degrees of freedom. For perturbative
gravity, it has been shown that \cite{schl}, the physical action is indeed positive definite. For non-perturbative gravity, this
gauge fixing is non-trivial, and it is difficult to find the `physical degrees of freedom'. 

A clear procedure however exists for identifying the physical measure, using the Faddeev-Popov determinant. We use this and perform a non-perturbative calculation, to identify the `physical action' for the trace sector of the theory. 
A generic metric can be written as
\be
g_{\m \n}= e^{2 \phi}\bar g_{\m \n}= e^{2\phi}\frac{\partial X^{\l}}{\partial x^{\m}}\frac{\partial X^{\r}}{\partial x^{\n}} g^{\perp}_{\l \r}
\ee

where $g^{\perp}_{\m \n}$ is a gauge fixed metric, and $X^{\mu}(x^\m)$ is a diffeomorphism and $\phi$ is fixed by the Yamabe condition. 

An infinitesimal version of this can be obtained in the cotangent space of the metric space as
\be
h_{\mu \nu} = h^{\perp}_{\mu \nu} + (L\xi)_{\mu \nu} + \left(2\tilde\phi + \frac12 \nabla\xi\right)g_{\mu \nu}
\label{coord}
\ee
Here $h^{\perp}_{\mu \nu}$ is the traceless part of the gauge invariant metric, $\xi_{\mu}$ the generator of diffeomorphisms, and the 
\be
(L\xi)_{\mu \nu} = \nabla_{\mu}\xi_{\nu} + \nabla_{\nu}\xi_{\mu} - \frac12 \nabla.\xi \ g_{\m \n}
\label{tr}
\ee
is an operator which maps vectors to traceless tensors. The last term of (\ref{coord}) corresponds to the trace sector of the metric including pure conformal orbits generated by $\tilde\phi$ representing an infinitesimal change in the conformal mode.

This coordinate transformation (\ref{coord}), when implemented in the measure in cotangent space leads to a Jacobian,
the Faddeev-Popov determinant. The same determinant can be used for the measure in the path-integral,
where a coordinate transformation to the physical coordinates $g_{\m \n}^{\perp}$ and conformal mode $\phi$, and the diffeomorphisms $\xi_{\m}$
is implemented \cite{bbm}. In this paper we explicitly isolate the contribution of the trace part of the diffeomorphisms to the Faddeev-Popov
determinant by parametrising the trace by a scalar field. This is described next.

\subsection{Redefining Diffeomorphisms}

The vector $\xi$ which generates the diffeomorphism is broken up into a divergence less vector $\hat \xi$ and a divergence of a `scalar' $\sigma$
\be
\xi_{\mu}= \hat\xi_{\mu} + \nabla_{\mu}\sigma
\label{decomp}
\ee
\be
\nabla^{\mu}\xi_{\mu} =  \nabla^{\m}\nabla_{\m}\sigma
\label{scale}
\ee

Thus the trace of the diffeomorphism $\nabla.\xi$ is a function of the scalar $\sigma$.
One more aspect of this discussion is that one can write $\hat \xi_{\mu}$ in terms of the divergence
of an antisymmetric two tensor (this is discussed in \cite{adg2})

The equation (\ref{scale}) determines $\sigma$ up to a scalar whose divergence is zero,
 in terms of the trace of the diffeomorphism. 
The very interesting aspect of this breakdown is the fact that `orthogonal' orbits of $\sigma$ which do not contribute to the traceless part of the diffeomorphisms are
not the `conformal killing' orbits, but solutions to the equation
\be
\left(\nabla_{\mu}\nabla_{\nu}-\frac12 g_{\mu \nu}\nabla^2\right)\sigma=0
\label{vanish}
\ee
This does not reduce to the equation for solving for conformal killing vectors, 
\be
(L\xi)_{\m \n}=0
\ee 
which is a much restricted equation for $\xi_{\m}$. 

\section{The Measure in the Path-Integral}
We now calculate the Faddeev-Popov determinant, but using (\ref{decomp}).
The interesting
aspect of this new calculation is the separation of the trace of the diffeomorphisms and evaluation
of the Faddeev-Popov determinant for this separately. 
The path-integral is defined to be
\be
Z= \int {\cal D}\phi~ {\cal D} \bar g_{\mu \nu} \exp\left(\frac1{16\pi G}\int d^4 x e^{2\phi}~\sqrt{\bar g} [\bar R + 6 (\bar\nabla\phi)^2]\right)
\ee
(we subsequently set $16\pi G$=1)
To identify the
correct measure one writes in the cotangent space of the De-Witt super space
\cite{bbm} or Equation (\ref{coord}), a coordinate transformation to the
traceless gauge fixed part (denoted by a $\perp$) and a trace part and the pure diffeomorphisms
generated by
$\xi_{\mu} = {\hat \xi}_{\mu} + \nabla_{\mu} \sigma$.
The Jacobian of the coordinate transformations is the Faddeev-Popov determinant
and written as (${\rm det M}$) subsequently.
\be
Z= \int{\cal D}\phi ~{\cal D} g^{\perp}_{\mu \nu}~{\cal D} {\hat \xi}_{\mu}~ {\cal D}{\sigma} ~{\rm det M}~ \exp\left( \int d^4x \sqrt{\bar g}~ e^{2 \phi}[\bar R + 6(\bar\nabla \phi)^2]\right)
\ee
This way of gauge fixing is completely non-perturbative, and is not specific to a gauge fixing. The {\rm det M} is shown to
have a scalar determinant (Appendix A) times a vector and a tensor determinant.
\be
{\rm det M} = {\rm det}_S [ 8 (1+2C)\left(-2(\nabla)^4 + 4\nabla_{\mu}\nabla^2\nabla^{\mu} + 4 \nabla_{\mu}\nabla_{\rho}\nabla^{\mu}\nabla^{\rho}\right)]^{1/2} {\rm det_V \tilde V}{\rm det_T T}
\label{scalar}
\ee
The operators $V_{\mu}$ and $T_{\mu \nu}$ which are vector and tensor operators can be easily identified as given in the appendix. 
$\nabla^{\mu}$ is the covariant derivative operator and $\nabla^2$ is the Laplacian with respect to the metric $g_{\m \n}$ (In case the derivative operators are evaluated 
for $\bar g_{\m \n}$ they are represented by $\bar\nabla$ operators). $C$ is the constant in the De-Witt metric
which determines the signature (see Appendix A for definition). The De-Witt metric is of indefinite signature for $C<-1/2$. 
The gauge volume or the integrations over the ${\hat \xi}_{\mu}$ and the $\sigma$ can be taken out of the path-integral, and the
Faddeev-Popov determinant remains to contribute to the effective action. 
Note that since this isolation of the trace is independent of the gauge fixing
condition, our results will be true in any gauge.
\section{The resolution of the conformal mode problem}
\subsection{Perturbative Case} 
We begin by taking the perturbative case and give the resolution of the
conformal mode problem in some known cases using the measure in (\ref{scalar}).
In the perturbative situation $\phi$ is taken to be very small, thus $e^{2\phi}= 1+ 2\phi + 2\phi^2$ and $\bar R$ is the curvature
of the space-time one is perturbing about; hence
the action gives
\bea
S & = & -\int d^4x \sqrt{\bar g} (1+ 2 \phi + 2 \phi^2) [\bar R + 6 (\bar\nabla\phi)^2]\\
&=& -\int d^4 x \sqrt{\bar g} [6 (\bar\nabla \phi)^2 + 2 \phi \bar R + 2\phi^2 \bar R + \bar R]\\
&=&-\int d^4 x \sqrt{\bar g} \left\{2\phi'[-3\bar \nabla^2 +  \bar R]\phi' + \bar R - \frac{1}{2}\bar R (\Delta^{-1})\bar R\right\}
\eea
Where the square in $\phi'= \phi + \frac12 \Delta^{-1} \bar R$ is completed and one obtains a non-local term in that process ($\Delta\equiv - 3\bar\nabla^2 + \bar R$).

For perturbations about Minkowski space-time,
 $R_{\mu \nu \lambda \rho}=0$ in the first approximation, and one obtains
the scalar determinant (\ref{scalar}) as
\be
\rm det_S [8 (1+2C)(6\bar \nabla^4)]^{1/2}
\ee
Under suitable boundary conditions, the determinant splits into 
\be
\rm det_{S}[-8 (1+2C) 6\bar \nabla^2]\rm det_{S}[-\bar \nabla^2]
\ee
The way we spilt the fourth order determinant, one corresponds to a convergent determinant ${\rm det}_s (-\bar\nabla^2)$, the other operator
has divergent determinant for $C< -1/2$. The zero modes of the above 
determinants and the conformal killing directions, have to be factored to give a meaningful answer \cite{mm}. For the purposes
of this paper, we assume that they have been factored as required. 
Thus the partition function acquires the following form in the conformal sector:
\be
\int {\cal D \phi} \ \rm det_{S}[-48(1+2C)\bar \nabla^2]^{1/2} \rm det_S[-\bar \nabla^2]^{1/2} exp(-\int d^4 x \ \sqrt{g} \ \phi'[6\bar \nabla^2]\phi')
\ee
Clearly the $\phi$ integral can be written formally as a determinant, which is divergent. 
\be
\rm det_{S}[-48(1+2C)\bar\nabla^2]^{1/2}\rm det_S[-\bar \nabla^2]^{1/2} \frac{1}{\rm det_S [6\bar \nabla^2]^{1/2}}
\ee
Thus the divergent determinants from the Faddeev-Popov and the $\phi$ integral can be cancelled for $C<-1/2$. The remaining terms after
the cancellation are convergent and will give finite answers for the path-integral.

Next the de-sitter and anti de sitter backgrounds which are constant
curvature metrics and hence have $\bar R_{\mu \nu}= \Lambda g_{\mu \nu}$ are considered ($\Lambda$ is a cosmological constant), and the
conformal mode is treated as a fluctuation over that.

The scalar determinant (\ref{scalar}) factorises rather neatly as fourth order operator is
\be
6\nabla^4 +4 [\nabla_{\mu},\nabla^2]\nabla^{\mu} + 4 \nabla_{\mu}[\nabla_{\rho},\nabla^{\mu}]\nabla^{\rho}= 6\nabla^4 + 8\Lambda\nabla^2
\ee
The determinant can be factorised for the above operator using $\bar R=4 \Lambda$, as
\be
\rm det_{S}(-\bar \nabla^2)[8(1+2 C)(-3\bar \nabla^2- \bar R)]= \rm det_{S}(-\bar \nabla^2)\rm det_S[8(1+2C)(-3\bar \nabla^2 - \bar R)].
\label{factor}
\ee
The action for the scale factor in the de-sitter or anti-desitter backgrounds is

\be
\int d^4 x  \sqrt{\bar g} (1+ 2\phi + 2\phi^2) \left[\bar R + 6(\bar \nabla\phi)^2\right] + \int d^4 x \sqrt{\bar g} (1+ 4\phi + 8 \phi^2) (-2\Lambda)
\ee
When the square is completed one obtains $\int d^4 x \ \sqrt{\bar g} \ 2 \phi'(-3\bar \nabla^2 - \bar R)\phi' $
 where we have put in $\bar R=4 \Lambda$. The integral over $\phi'$ thus gives a divergent determinant which is inverse to second of the
factored determinants of (\ref{factor}) and is thus cancelled. Thus the partition function for pure conformal fluctuations about de Sitter space
is given by
\be
N [{\rm det}_s(-\bar \nabla^2)]^{1/2}
\ee
Where N includes normalisation, but clearly the conformal mode divergence has been cancelled.
So, for most Ricci flat and constant curvature metrics, we seem to have
correctly identified the resolution. This resolution is similar to that described in \cite{mm}. 
There remain the transverse fluctuations of the metric, however which is beyond the scope of the discussion of the conformal mode.
\subsection{Non-perturbative Case}
For the non-perturbative case, no such neat factorisations of the determinant occur, and one has to compute the functional determinant using known techniques like the heat kernel equation. Ab initio, the evaluation of the determinant (\ref{scalar}) in the non-perturbative regime appears very difficult, as the operator is a fourth order differential operator.
The scalar operator (\ref{scalar}) is clearly a self adjoint operator (the $\nabla^4$ term is obviously self adjoint and the $\nabla_{\mu}R_{\mu \nu}\nabla_{\nu}$
is also self adjoint due to the symmetry of $R_{\mu \nu}$ under the interchange of the $\mu$ and the $\nu$ indices).
But we can try to obtain the determinant using Heat Kernel techniques \cite{dewit,avramidi,full}. 

The heat kernel is defined in order to achieve a $\zeta$ function regularisation of the determinant. Given an operator $F$ with eigenvalues
$\lambda$, the zeta function is defined to be
\be
\zeta(p)= \sum_{\lambda}\frac1{\lambda^{p}}
\ee
Clearly, the determinant of the operator, would be given by

\be
\rm det(F)= e^{\rm Tr ln F}
\ee
 
and thus

\be
\rm det(F) = e^{-\zeta'(0)}.
\label{zeta}
\ee
 
Other regularisation schemes can also be used
if required, thus the exact answer would be particular to the way of regularisation. From (\ref{zeta}), the Faddeev-Popov determinant can be written as
an exponential and adds to the classical action in the `weight' of the path-integral creating an effective action. Since the Faddeev-Popov has the 
square root of the determinant appearing in the measure, the exact terms which appear in the effective action for the scalar determinant
is $\G_{\rm trace}= \frac12 \zeta'(0)$. Thus one has to find $\zeta'(0)$ for the scalar determinant of (\ref{scalar}).

The zeta function for a given operator is appropriately written in terms of the `Heat Kernel'.  
\be
\zeta(p)= \frac{\mu^{2 p}}{\Gamma(p)} \int dt\  t^{p-1} {\rm Tr} \exp(-t F)(x,x') 
\ee
The Heat Kernel is precisely the term $U(t, x,x')=\exp(-t F)(x,x')$, $t$ is a parameter and $x,x'$ represent coordinates of the manifold in which the operator is defined. The heat kernel also satisfies the differential equation
\be
\left(\frac{\partial}{\partial t} + F\right) U(t, x,x')= \delta(x,x')|_{t=0}
\label{adg}
\ee
where the boundary condition is that of a diffusion equation. The Heat Kernel can be solved exactly for the Laplacian in flat space,
and for curved space-time, De-Witt wrote a particular ansatz, which has been solved partially.
The ansatz for the Heat Kernel for the Laplacian is an expansion in powers of the parameter $t$
\be
U(t,x,x')= \frac1{(4\pi t)^2}e^{-\frac{\bar\sigma}{2t}} \sum_{n} a_n (x,x') t^n
\ee
where $\bar\sigma$ is half of the square of the geodesic distance existing given two points (x,x'). The actual evaluation of the
heat kernel and the determinant of the operator is done in the coincidence limit $x\rightarrow x'$.

The procedure for finding the heat kernel for the scalar operator which appears in the determinant in the
measure is simplified slightly, by solving in two different regimes. 
The scalar operator in (\ref{scalar}) is taken thus
\be
-2 \nabla^4 + 4\nabla_{\mu}\nabla^2\nabla^{\mu} + 4 \nabla^{\mu}\nabla^{\rho}\nabla_{\mu}\nabla_{\rho}
\ee
By using the commutation relations, one obtains
\bea
&&6\nabla^4 + 4\nabla_{\mu}[\nabla^2,\nabla^{\mu}] + 4 \nabla^{\mu}[\nabla^{\rho},\nabla_{\mu}]\nabla_{\rho}\\
&&6 \nabla^4 + 8 \nabla_{\mu} R^{\mu \nu}\nabla^{\nu} \label{rew}
\eea

We take two limits, one where there is weak field gravity, and one obtains $\nabla_{\m} R^{\mu \nu}\sim 0$, here (\ref{rew})
is approximated by
\be
6\nabla^4\left(1+ \frac{4}{3}(\nabla^4)^{-1}\nabla_{\mu}R^{\mu \nu}\nabla_{\nu}\right).
\ee
And where gravity is strong, one obtains
\be
8\nabla_{\mu}R^{\mu \nu}\nabla_{\nu}\left(1+ \frac{3}{4}(\nabla_{\mu}R^{\mu \nu}\nabla_{\nu})^{-1}\nabla^4\right).
\ee
{\it This particular approximation does not appear in any previous calculation, usually the expansion of higher order operators has been 
done in orders of $(\nabla^2)^{-1}$ \cite{bar}.}
Thus for the purposes of the kinetic term of the conformal mode, which is the reason for the unboundedness
of the Euclidean action, it is enough to obtain the determinant of the following scalar operators in the above two
limits. 
\be
\rm det_S \left(8(1+2C) 6\nabla^4\right) \ \ \ \ \ (\nabla^{\m}R_{\m \n} \sim 0)
\label{operator1}
\ee
and
\be
\rm det_S\left(8(1+2C) 8 \nabla_{\mu}R^{\mu \nu}\nabla_{\nu}\right) \ \ \ \ \ \ \ (\nabla^{\m}R_{\m \n}\gg 1)
\label{operator}
\ee

The first one (\ref{operator1}) can be evaluated using the Heat Kernel for the Laplacian for arbitrary space times, under certain
boundary conditions, by splitting the fourth order operator into product of two Laplacians whose heat kernel expansions
are well known \cite{dewit}.

 The (\ref{operator}) operator is clearly non-Laplacian type and non-minimal in the sense that $R_{\m \n}$ is not covariantly constant.
Thus, I use a new ansatz for the Heat Kernel, where
\be
U(t,x,x')= \frac1{(4\pi t)^2}\exp(-\ve/2t)\sum_{n} a_n t^n
\ee
where $\ve$, is a generalisation of $\bar\sigma$ used by De-Witt, and is determined by solution to an equation written below. 
One then finds a recursion relation for the $a_i$ using the differential equation for the definition of the Heat Kernel (\ref{adg}).

The heat kernel differential equation for $t\neq 0$ gives for the operator $(- 8\nabla^{\m}R_{\m \n}\nabla^{\n})$
\bea
\left(\frac{\partial}{\partial t} -8 \nabla_{\m}R^{\m \n}\nabla_{\n}\right)\frac{1}{(4\pi t)^2}\exp(-\ve/2t)\sum a_n t^n&=&0\\
{\rm or} \ \ \left(\frac{\partial}{\partial t} - 8\nabla^{\m}R^{\m \n}\nabla^{\n} - 8 R^{\m \n}\nabla^{\m}\nabla^{\n}\right)\frac{1}{(4 \pi t)^2}\exp(-\ve/2t)\sum a_n t^n =0
\eea
The linear term can be removed by a scaling by $\exp\left(\Lambda(x)\right)$. The function $\Lambda$ is determined 
according to the differential equation $\nabla_{\tau} (\Lambda)= -\frac12 (R^{-1})_{\n \t}\nabla_{\l} R^{\l \n}$. This reduces the above equation to
\be
\left(\frac{\partial}{\partial t} - 8 R_{\m \n} \nabla^{\m}\nabla^{\n} + Q \right)\frac1{(4 \pi t)^2}\exp(-\ve/2t)\sum a_n t^n=0
\label{new}
\ee
where a potential $Q= -8 \nabla_{\mu} R^{\mu \nu} \nabla_{\n}\L - 8 R^{\m \n} \nabla_{\m} \nabla_{\n} \L$ gets added due to the scaling of the wavefunction.
The recursion relations for coefficients $a_n$ obtained from (\ref{new}) are:
\bea
\frac{\ve}{2} - 2 R_{\m \n} \nabla^{\mu}\ve \nabla^{\nu}\ve &= &0 \label{ve}\\
8 R_{\m \n}\nabla^{\mu}\nabla^{\nu}a_{n-1} - 8 R_{\m \n} \nabla^{\nu}\ve \nabla^{\mu} a_{n} - 4 \nabla^{\mu}\nabla^{\nu}\ve R_{\m \n} a_{n}
- (n-2)a_{n} - Q a_{n-1}=0
\label{split}
\eea

The obvious solution for (\ref{ve}) is to take $\frac14\ve= R_{\m \n} \nabla^{\mu}\ve \nabla^{\nu}\ve$.
This equation is solved easily in the case where the Ricci curvature is constant $R_{\mu \nu} = \frac{1}{l^2} g_{\m \nu}$, and using that a generalisation is given for arbitrary curvature metrics
in Riemann normal coordinates.

The case of the constant curvature (\ref{new}) is solved by:
\bea
\frac{\ve}{4} & = & \frac1{l^2}g_{\mu \nu}\nabla^{\mu}\ve\nabla^{\nu}\ve \\
\ve & =  &  \frac{l^2}{8}\bar\sigma
\label{eq:const}
\eea

where $\bar\sigma$ is half of the square of the geodesic distance from the point $x$ to $x'$.

The case of arbitrary curvature, the Riemann normal coordinate expansion is taken for the metric and a solution is given in the coincidence
limit only $x\rightarrow x'$. The Riemann normal coordinates are used to expand for the metric and the curvature, about a given point, similar to
a Taylor series for a function. The standard form for the square of the geodesic distances is
\be
2\bar \sigma(x, x')= g_{\alpha \beta}\bar\sigma^{\alpha}\bar\sigma^{\beta} 
\ee

where $\bar\sigma^{\alpha}(x,x')$ is the tangent vector to the geodesic at the point (x). In Riemann Normal coordinates, this 
is
\be
2\bar\sigma (x, x')= \d_{a b} \hat x^a \hat x^b
\ee
where $\hat x^a$ is the coordinate joining the origin of the Riemann normal coordinates to a nearby point, or $(x-x')^a$ when $x$ is taken
as the origin and $x'\rightarrow x$.
Motivated from equation (\ref{eq:const}), the $\ve$ can be written in Riemann normal coordinates as:
\be
\ve= \frac1{16}g_{ ab} (R^{-1/2})^{ a c}\hat x_{c} (R^{-1/2})^{b d} \hat x_{d} 
\ee

This has the interesting properties in the coincidence limit (though there are no obvious interpretations for $\ve$ as there are for $\sigma$ in terms of geodesics
and tangents at the points x, x')
\bea
\nabla^{a}\ve  & = & 0  \label{def1} \\
\nabla^{a}\nabla^{b}\ve  &= & \frac{1}{8}(R^{-1})^{ab} \label{def2}\\
\nabla^{a}\nabla^{b}\nabla^{c}\ve &=& 0 \label{def3} \\
\nabla^{a}\nabla^{b}\nabla^{c}\nabla^{d}\ve  & = & -\frac12 R^{a}{\ _e} \ ^{b} {\ _f} (R^{-1/2})^{ec} (R^{-1/2})^{fd} \label{def4}
\eea 
$(R^{-1/2})^{ac}$ is the square root of the inverse matrix of $R_{ac}$ at the origin of the Riemann normal coordinates.

In the coincidence limit, one finds that, the recursion relation  obtained in (\ref{split}) reduces to
\be
n~ a_n - 8 R_{\m \n}\nabla^{\mu}\nabla^{\nu}a_{n-1} + Q a_{n-1}=0
\label{recur}
\ee

To solve this recursion relation, we act on the 
split equation (\ref{split}) again with the operator $R_{\m \n}\nabla^{\m}\nabla^{\n}$, and 
one gets a equation for $8 R_{a b} \nabla^{a}\nabla^{b} a_n$ which
is solved and substituted in (\ref{recur}).
The coefficients $a_o$ and $a_1$ are solved here. 
 
As per the boundary condition, $a_0=1$. In addition, we get the the derivatives of $a_0$ from the coincidence
limit of the equations in Appendix B.
\bea
\nabla^{a}a_0 &= &\frac12 (R^{-1})^{cd}\nabla^a R_{cd} \label{def5}\\
R_{a b}\nabla^{a}\nabla^{b}a_0&=& -\frac{1}{4}R_{c d}[(R^{-1})^{a b } \nabla^{c} R_{a b}] [(R^{-1})^{a'b'}\nabla^{d} R_{a' b'}] \nn \\
&-& \frac12 R_{c d} (R^{-1})^{a b}\nabla^{c}\nabla^{d}R_{a b} -R^2 \label{def6}
\label{eq:solve}
\eea
(In the coincidence limit the  indices of the Riemann Normal coordinates can be interchanged with the $\mu$ indices). These equations are used in equation (\ref{recur}) to get $a_1$. Similarly a $a_n$ for arbitrary $n$ can be obtained.
The equation for $a_1$ is
\bea
a_1= 8 R_{\m \n}\nabla^{\m}\nabla^{\n} a_0 - Q a_0
\label{coeff}
\eea
Substituting for $Q$ and from (\ref{eq:solve}), we get an expression for $a_1$ in terms of the curvature
invariants and the derivatives of the curvature. This is thus a derivation of the Heat Kernel expansion
of a non-Laplacian operator for non-perturbative gravity. In this derivation of the heat kernel expansion, we have ignored the boundary
of the manifold, as we have done a Riemann Normal coordinate expansion about one local point. Inclusion of half-integer $t$ coefficients ensures the
inclusion of boundary terms, but that is not relevant for the calculation in this paper. 
\subsection{The non-perturbative resolution} 
In the two regimes we considered, $\nabla^{\m}R_{\m \n} \sim 0$ and $\nabla^{\m}R_{\m \n} \gg 1$, the terms
in the $\G_{\rm trace}$ or $\frac12 \zeta'(0)$ which will add to the $-\int d^4 x e^{2\phi} \sqrt{\bar g}(\bar\nabla \phi)^2$ term in the classical action is the term with $a_1$ coefficient. The $a_2$ and higher coefficients are
higher derivative terms and are not relevant for the discussion. So we will isolate the $a_1$ 
term in the zeta function's derivative, and see why we are convinced that the negative term in the classical action
is taken care of by the measure's $\G_{\rm trace}$.
The zeta function and its derivative are: 
\bea
\zeta (p) &= & \frac{\mu^{2p}}{16 \pi^2 \Gamma(p)} \int dt \ t^{p-3} \ {\rm Tr}_{x} e^{-t Q} \ \sum_{i}  a_i t^i \\
16 \pi^2\zeta'(p) &=& \frac{\mu^{2 p}}{\Gamma(p)} \ \left(\ln \mu^2 - \frac{\Gamma'(p)}{\G(p)}\right) \ \int dt t^{p-3} \ {\rm Tr}_x e^{-t Q} \ \sum_{i}  a_i t^i \nn \\
&+& \frac{\mu^{2p}}{\G(p)} \ \int dt \ t^{p-3} \ \ln t \ {\rm Tr}_x \ e^{-t Q} \ \sum_i a_i t^i 
\label{eq:deriv}
\eea

The finite term as $p\rightarrow 0$ appears in the last term of the derivative of the zeta function (as we know
this might not be the unique the way to extract the finite term but sure explains a way to cancel the negative
term from the classical action). This is given in details in the Appendix. The finite term (regulator independent) is remarkably proportional to $a_1$  and is obtained from (\ref{eq:deriv}) in the
Appendix as
(using a regularisation $\zeta(0)=-1/2$.)
\be
\zeta'(0)_{\rm finite}= -\frac{1}{32 \pi^2} {\rm Tr}_x a_1 
\ee
 
Thus in the effective action $ \G_{\rm trace} + \G_{\rm classical}= \frac12 \zeta'(0) + \G_{\rm classical}$ we get to first order $-\frac{1}{64 \pi^2}{\rm Tr}_x a_1 + \G_{\rm classical}$. In the subsequent
discussion, we fix the $a_1$ in the weak gravity and strong gravity regimes and find the $\G_{\rm trace}+ \G_{\rm classical}$.
 
(i) $\nabla_{\mu}R_{\m \n}\approx 0$, the Faddeev-Popov determinant is 
\be
{\rm det}(8(1+2C)6\nabla^4)= {\rm det}(-8(1+2C)6\nabla^2) {\rm det} (-\nabla^2)
\ee

In the factorised form, the first scalar determinant is a divergent one for $C<-1/2$, and the second one is a convergent one. We discuss the divergent determinant's
contribution to the effective action as this should cancel the divergence from the classical action.

The first scalar determinant is of a Laplacian and thus we use the heat kernel of a Laplacian, and analytically continue to the
divergent regime of $C<-1/2$.
The $a_1$
coefficient of the Laplacian is well known, and to quote \cite{dewit,avramidi}
\be
a_1= \frac{1}{6} R
\ee

I scale the coefficient $a_1$ by Planck length squared ($ G h/2\pi$) to get the exponential dimensionless and restore the $16\pi G$ in the classical action. Writing $R$ in terms of $\bar R$ and $(\nabla \phi)^2$, and using the constants in the determinant one gets
as the coefficient of the kinetic term of the conformal mode in the effective action ($\G_{\rm trace} + \G_{\rm classical}$)
\be
-\frac{1}{16 \pi}\left[1 + \frac{2(1+ 2C)}{\pi}\right]
\ee

Thus the positive action takes over at $(1+2C)> -  \pi/2$. From Einstein' action $C=-2$, and Euclidean Einstein gravity has a positive definite effective action. The number $-\pi/2$ might differ for different
regularisation schemes, (and conventions for defining determinants from Gaussian integrals) but it is indeed a finite number, and we should be in the realm of a
convergent path-integral for Euclidean quantum gravity. Note that the effect of this calculation has been in the end change of the
overall sign of the action by a minus sign. This is what has been observed in the continuum limit of the discrete lattice gravity calculation
of the path-integral by Ambjorn, Jurkiewicz and Loll \cite{loll}. So, these calculations confirm their observations of the effective action.

(ii) In the case of the regime $\nabla^{\mu}R_{\m \n} \gg 1$, the relevant operator is (\ref{operator}) and the coefficient $a_1$ is from (\ref{coeff}, \ref{eq:solve}) (The Q in (\ref{coeff}) is absorbed
in the exponential of (\ref{eq:deriv})),
\bea
a_1& =& -2 R_{\l \s}[(R^{-1})^{\m \n} \nabla^{\l} R_{\m \n}] [(R^{-1})^{\m'\n'}\nabla^{\s} R_{\m' \n'}] \nn \\ 
&-& 4 R_{\l \s} (R^{-1})^{\m \n}\nabla^{\s}\nabla^{\l}R_{\m \n} - 8 R^2
\label{a12}
\eea
We find that 
  $-R^2$ term has exactly the
sign required to cancel the contribution from the $(\bar \nabla\phi)^2$ term in the classical action, as writing R in terms of $\bar R$ and $\bar \nabla \phi$, one finds $(\bar\nabla\phi)^4$ from $R^2$,
which dominate for configurations with rapidly varying $\phi$. The other terms do not give divergent negative terms. Thus $\G_{\rm trace} + \G_{\rm classical}$
in this regime also emerges as positive definite. 
Note in this non-perturbative regime, the effective action at this order does not merely change by an overall minus sign but has additional non-trivial contributions
proportional to ${\bar R}^2$ and higher derivative terms.

\section{Discussions}
In this paper, I isolated the scalar determinant in the Faddeev-Popov measure of the gravitational path integral and computed its contribution to the
effective action using heat kernel techniques. It was found that this contribution added to the negative
classical action to render it positive. In the process, I found the heat kernel of an operator $R_{\m \n}\nabla^{\m}\nabla^{\n}$, in the non-perturbative
regime, and this is a useful result.

{\bf Acknowledgement:} I would like to thank T. Budd for useful comments and pointing out typos in equation (\ref{split}).

\appendix{Appendix A: Derivation of the Faddeev-Popov determinant}\\

To derive the Faddeev-Popov determinant, for the particular gauge fixing of the gravitational path-integral,
we use the method described in \cite{mm, bbm}. The Gaussian normalisation condition fixes any ambiguity in the
coefficients completely. We take the diffeomorphism transformations of the metric cotangent space elements
and the new vector and scalar fields by
\be
h_{\m \n}= h_{\m \n}^{\perp} + (L (\hat\xi,\sigma))_{\mu \nu} + (2\bar\phi + \frac12 \nabla^2 \sigma)g_{\mu \nu}
\ee
Where 
\be
L(\hat \xi,\phi)_{\mu\nu}= \nabla_{\n}\hat\xi_{\mu} + \nabla_{\m}\hat\xi_{\nu} + \left(2 \nabla_{\mu}\nabla_{\nu} - \frac12 g_{\m \n} \nabla^2\right)\sigma
\ee
The Gaussian normalisation condition states that
\be
1= \int {\cal D}h_{\m \n} \exp{(- \int \sqrt{g} ~d^4 x \ h_{\m \n}~G^{\m \n \r \t}~h_{\r \t})}
\ee
where $G^{\m \n \r \t}$ is the DeWitt supermetric obtained in terms of the background metric as
\be
G^{\m \n \r \t}= \frac12 \left(g^{\m\r}g^{\n \t}+g^{\m \t}g^{\n \r} + C g^{\m \n}g^{\r \t}\right)
\ee

Interchanging the coordinates of the tangent space leads to the determination of a Jacobian (a function of the metric), 
which is then determined as, 
\be
J^{-1}= \int {\cal D} h^{\perp}_{\m \n}{\cal D}\sigma{\cal D} \hat\xi_{\mu} {\cal D}\bar\phi \exp \left(- \int~ d^4x ~\sqrt{g} \ h_{\m \n}~ G^{\m \n \r \t}~ h_{\r \t}\right) 
\ee

The scalar product in the exponent breaks up into
\bea
 &= & \int d^4 x~\sqrt{g}~ {h^{\perp}_{\m \n}G^{\m \n \r \t} h^{\perp}_{\r \t}} + \int \sqrt{g} \ d^4x \ {L_{\m \n}G^{\m \n \r \t} L_{\r \t }}
+ \int d^4x \ \sqrt {g}~ 2~h^{\perp}_{\m \n}G^{\m \n \r \t}L_{\r \t}
\nn \\ &+ & 8(1+2 C)\int  d^4 x ~\sqrt{g}~ \Omega^2 
\eea

where $\Omega= \bar \phi+\frac14\nabla^2\sigma$, and can be just integrated as a redefinition of the
conformal mode $\bar \phi$. Thus the Faddeev-Popov determinant gets only a factor of constant.
The interesting terms are contained in 
\bea
\int d^4x ~\sqrt{g} ~ L_{\m \n}G^{\m \n \r \t}L_{\r \t}&= &\int d^4 x \sqrt{g}\left[ \left\{\nabla_{\n}\hat\xi_{\m}+ \nabla_{\m}\hat \xi_{\n}
+ (2\nabla_\m\nabla_\n -\frac12\nabla^2 g_{\m \n})\sigma) \right\} \right.  \nn \\& \times & \left.G^{\m \n \r \t}\left\{\nabla_{\rho}\hat\xi_{\t}+\nabla_{\t}\hat\xi_{\rho} + (2\nabla_{\r}\nabla_{\t} - \frac12 g_{\r \t}\nabla^2)\sigma\right\}\right]
\eea
These terms give the Faddeev-Popov determinant from the vector and the scalar
completion of squares.
The completion of the square in the scalar sector gives this term:
\be
\int d^4x \ \sqrt{g} \ \sigma' \left(2\nabla_{\m}\nabla_{\n} -\frac12 g_{\m \n} \nabla^2\right)G^{\m \n \t \s}\left(2\nabla_{\t}\nabla_{\s} - \frac12 g_{\t \s} \nabla^2\right)\sigma'
\label{scal1}
\ee
The simplification of which gives 
\be
\int d^4 x \ \sqrt{g} \  \sigma' \left[- 2 (\nabla)^4 + 4 \nabla_{\mu}\nabla^2 \nabla^{\mu} + 4\nabla_{\m}\nabla_{\r}\nabla^{\m}\nabla^{\r}\right]\sigma'
\label{scal}
\ee
where the $\sigma'= \sigma + X_{\rho}\hat \xi^{\rho} + Y^{\m \n} h^{\perp}_{\m \n}$ and
$X_{\rho}= \Delta ^{-1} \left[3\nabla_{\rho} \nabla^2 + R^{\t}_{\r} \nabla_{\t}\right] $ and the operator
$Y_{\m \n} = \frac12 \Delta^{-1}\left[2\nabla_{\mu}\nabla_{\nu} -\frac12 \nabla^2 g_{\m \n}\right]$.
($\Delta$ is the scalar operator which appears in the squared term in (\ref{scal})).
The Integral is then completely Gaussian in each of the variables, and one obtains
\be
J^{-1}= \frac{1}{\sqrt{ 8(1+ 2C)}}\int {\cal D} h^{\perp}_{\m \n}{\cal D}\sigma {\cal D}\hat\xi_{\mu} \exp\left(-\int d^4x\ \sqrt{g} \  \left[h^{\perp}_{\m \n}T^{\m \n \t \r} h^{\perp}_{\t \r} + \hat \xi'_{\m} V^{\m \n} \hat \xi'_{\n} + \sigma'\Delta \sigma'\right]\right)
\ee
The integration of the above gives the
\be
J^{-1}= \frac{1}{\sqrt{ 8(1+ 2C) {\rm det_T}T {\rm det_V} V {\rm det_S} \Delta}}
\ee
 
Using the definition of determinants of operators. The operators $T$ and $V$ are tensor
and vector operators, and hence their determinants are tensorial and vectorial determinants.
We concentrate in obtaining the details of the scalar operator in this article.

\vspace{0.5cm}

\appendix{Appendix B: The $R_{\m \n}\nabla^{\m}\nabla^{\n}$ operator}\\
To find $\nabla^{\t}a_0$ and $R_{\t \s}\nabla^{\t}\nabla^{\s}a_0$ which appear in the defining relation for $a_1$, one 
takes the (\ref{recur}), and
operates on it further with $R_{\m \n}\nabla^{\m}\nabla^{\n}$.

\be
R_{\l \s}\nabla^{\l}\nabla^{\s}~ \left[(n-2)a_n + 4 \nabla^{\m}\nabla^{\n}\ve R_{\m \n} a_n + 8 \nabla^{\mu}\ve\nabla^{\nu} a_n R_{\m \n} 
-8 R_{\m \n} \nabla^{\m}\nabla^{\n} a_{n-1} + Q a_{n-1}\right] = 0 
\ee
which results in
\bea
(n-2) R_{\l \s}\nabla^{\l}\nabla^{\s} a_n && \label{eq:res}\\+ 4 R_{\l \s}\nabla^{\l}\nabla^{\s}\nabla^{\m}\nabla^{\nu}\ve R_{\m \n} a_n 
+ 4 R_{\l \s}\nabla^{\s}\nabla^{\mu}\nabla^{\nu}\ve \nabla^{\lambda}R_{\mu \nu} a_n + 4 R_{\l \s}\nabla^{\s}\nabla^{\mu}\nabla^{\nu}\ve R_{\m \n} \nabla^{\l}a_n && \nn \\
+ 4 R_{\l \s} \nabla^{\l}\nabla^{\m}\nabla^{\n}\ve \nabla^{\s} R_{\m \n} a_n + 4 R_{\l \s} \nabla^{\m}\nabla^{\nu}\ve \nabla^{\l}\ve\nabla^{\s}R_{\m \n} a_n
+4 R_{\l \s}\nabla^{\m}\nabla^{\n}\ve \nabla^{\s} R_{\m \n} \nabla^{\l} a_n  && \nn \\ + 4 R_{\l \s}\nabla^{\l}\nabla^{\m}\nabla^{\n}\ve R_{\m \n} \nabla^{\s} a_n
+ 4 R_{\l \s} \nabla^{\mu}\nabla^{\nu}\ve \nabla^{\lambda} R_{\m \n} \nabla^{\s} a_n
+4 R_{\l \s}\nabla^{\mu}\nabla^{\nu}\ve R_{\m \n} \nabla^{\l}\nabla^{\s}a_n && \nn \\ + 8 R_{\l \s}\nabla^{\l}\nabla^{\s}\nabla^{\m}\ve \nabla^{\n}a_n R_{\m \n} + 8 R_{\l \s}\nabla^{\s}\nabla^{\m}\ve\nabla^{\l}\nabla^{\n}a_n R_{\m \n}
+ 8 R_{\l \s}\nabla^{\s}\nabla^{\mu}\ve\nabla^{\nu}a_n\nabla^{\l}R_{\m \n} && \nn \\ + 8 R_{\l \s}\nabla^{\l}\nabla^{\m}\ve \nabla^{\s}\nabla^{\n} a_n R_{\m \n} + 8 R_{\l \s}\nabla^{\m}\ve\nabla^{\l}\nabla^{\n}a_n \nabla^{\s}R_{\m \n}
+ 8 R_{\l \s}\nabla^{\mu}\ve\nabla^{\s}\nabla^{\n}a_n \nabla^{\l}R_{\m \n} && \nn \\ + 8 R_{\l \s}\nabla^{\l}\nabla^{\m}\ve\nabla^{\n}a_n\nabla^{\s}R_{\m \n} + 8 R_{\l \s}\nabla^{\m}\ve \nabla^{\l}\nabla^{\n}a_n\nabla^{\s}R_{\m \n}
+8 R_{\l \s}\nabla^{\m}\ve\nabla^{\nu}a_n\nabla^{\l}\nabla^{\s}R_{\m \n} && \nn \\  - 8R_{\l \s}\nabla^{\l}\nabla^{\s}R_{\m \n}\nabla^{\m}\nabla^{\n}a_{n-1} - 8 R_{\l \s}\nabla^{\s}R_{\m \n}\nabla^{\l}\nabla^{\m}\nabla^{\n}a_{n-1}&&\nn\\
-8R_{\l \s}\nabla^{\l}R_{\m \n}\nabla^{\s}\nabla^{\m}\nabla^{\n}a_{n-1} -8 R_{\l \s}R_{\m \n}\nabla^{\l}\nabla^{\s}\nabla^{\m}\nabla^{\n}a_{n-1}+ 8R_{\l \s}\nabla^{\l}\nabla^{\s}Q a_{n-1} &&\nn \\
+8R_{\l \s}\nabla^{\sigma}Q\nabla^{\l}a_{n-1}+8R_{\l \s}\nabla^{\l}Q\nabla^{\s}a_{n-1}+8R_{\l \s}Q\nabla^{\l}\nabla^{\s}a_{n-1}&=&0 \nn
\eea
In the coincidence limit, apart from the terms containing $\nabla^{\m}\nabla^{\n} a_n$, there survives terms containing
one derivative of coefficients, to determine one takes one derivative of the recursion relation 
\bea
(n-2)\nabla^{\t}a_n + 4 \nabla^{\t}\nabla^{\m}\nabla^{\n}\ve R_{\m \n} a_n + 4 \nabla^{\m}\nabla^{\n}\ve\nabla^{\tau}R_{\m \n} a_n && \nn \\
+ 4\nabla^{\mu}\nabla^{\nu}\ve R_{\m \n} \nabla^{\tau} a_n +8 \nabla^{\t}\nabla^{\m}\ve\nabla^{\n}a_n R_{\m \n}
+8 \nabla^{\mu}\ve \nabla^{\t}\nabla^{\n}a_n R_{\m \n} && \\+ 8 \nabla^{\m}\ve\nabla^{\nu}a_n \nabla^{\tau}R_{\m \n}
-8\nabla^{\t}R_{\m \n}\nabla^{\m}\nabla^{\n}a_{n-1} - 8 R_{\m \n}\nabla^{\t}\nabla^{\m}\nabla^{\n}a_{n-1} +\nabla^{\t}Q a_{n-1}+Q\nabla^{\t}a_{n-1}&=&0 \nn
\eea 

In the coincidence limit, the non-zero terms from (\ref{eq:res}), are
\bea
(n-2)R_{\m \n} \nabla^{\m}\nabla^{\n} a_n + 4 R_{\l \s}\nabla^{\l}\nabla^{\s}\nabla^{\m}\nabla^{\n}\ve R_{\m \n} a_n
+4 R_{\l \s}\nabla^{\m}\nabla^{\n}\ve \nabla^{\l}\nabla^{\s}R_{\m \n} a_n &&\\
+4 R_{\l \s}\nabla^{\m}\nabla^{\n}\ve\nabla^{\s}R_{\m \n}\nabla^{\l}a_n 
+ 4 R_{\l \s}\nabla^{\m}\nabla^{\n}\ve\nabla^{\l}R_{\m \n}\nabla^{\s}a_n && \nn \\ + 4 R_{\l \s}\nabla^{\m}\nabla^{\n}\ve R_{\m \n}\nabla^{\s}\nabla^{\l}a_n+
8 R_{\l \s}\nabla^{\s}\nabla^{\m}\ve R_{\m \n}\nabla^{\l}\nabla^{\n}a_n + 8 R_{\l \s}\nabla^{\l}\nabla^{\m}\ve R_{\m \n}\nabla^{\s}\nabla^{\n}a_n && \nn \\
+ 8 R_{\l \s}\nabla^{\l}\nabla^{\n}\ve\nabla^{\s}R_{\m \n}\nabla^{\n}a_n- 8 R_{\l \s}\nabla^{\l}\nabla^{\s}R_{\m \n}\nabla^{\m}\nabla^{\n}a_{n-1}
-8 R_{\l \s}\nabla^{\s}R_{\m \n}\nabla^{\l}\nabla^{\m}\nabla^{\n}a_{n-1} && \nn \\ + 8R_{\l \s}\nabla^{\l}\nabla^{\s}Q a_{n-1} + 8 R_{\l \s}\nabla^{\l}Q \nabla^{\l}a_{n-1}
+8R_{\l \s}\nabla^{\l}Q \nabla^{\s}a_{n-1} + 8 R_{\l \s}Q \nabla^{\l}\nabla^{\s}a_{n-1}&=&0 \nn
\eea

Putting equations (\ref{def1},\ref{def2},\ref{def3},\ref{def4}) in the above one gets (\ref{def5},\ref{def6}).

\noindent
\appendix{Appendix C: The zeta function derivative}\\

As given in equation (\ref{eq:deriv}), the derivative of the zeta function is
\bea
16 \pi^2 \zeta'(p)&=& \frac{\m ^{2p}}{\G(p)} \ \left[ \ln \m^2 - \frac{\G'(p)}{\G(p)}\right] \int dt \ t^{p-3} \ {\rm Tr}_x \ e^{- Q t} \sum_i a_i t^i \nn \\
&+ & \frac{\m ^{2 p}}{\G(p)} \int dt \ t^{p-3} \ \ln t \ {\rm Tr}_x e^{-Q t} \sum_i a_i t^i 
\eea  
The last term gives the integral to be
\bea
&=& \frac{\m^{2p}}{\G(p)}\int dt \ t^{p-3} \ \ln t \ {\rm Tr}_x e^{-Q t} \sum_i a_i \ t^i \nn \\
&=& \frac{\m^{2p}}{\G(p)} \int dt \ t^{p-3} \sum_{n=1}^{\infty} \frac{(-1)^{n+1}}{n}(t-1)^{n} {\rm Tr}_x \ e^{-Q t} \sum_i a_i \ t^i \nn \\
&=& \frac{\m^{2p}}{\G(p)}\int dt \ t^{p-3} \sum_{n=1}^{\infty} \sum_{j=0}^n \frac{(-1)^{n+1}}{n} \ ^n C_j \ t^j \ (-1)^{n-j} \ {\rm Tr}_x \ e^{-Q t} \ \sum_i \ a_i \ t^i \nn\\ 
&=& \frac{\m^{2p}}{\G(p)} \sum_{i,j,n}\frac{(-1)^{2n+1-j}}{n} \Gamma(p-2+i+j) \  ^n C_j {\rm Tr}_x \ a_i \ Q^{2-p-i-j}
\eea

The term which is finite and non-zero in the above is obtained for $i=1$ and $j=1$, and plugging this in the derivative, and taking the limit $p\rightarrow 0$ is precisely
\be
\sum_{n=1}^{\infty} (1) {\rm Tr}_x \ a_1 = \zeta(0) {\rm Tr}_x \ a_1
\ee


\begin{thebibliography}{99}
\bibitem{hawk} G. W. Gibbons, S. W. Hawking, M. J. Perry, Nucl. Phys. {\bf B 138} 141 (1978).
\bibitem{schl} J. Hartle, K. Schleich, Phys. Rev. {\bf D 36} 2342 (1987).
\bibitem{schl1}K. Schleich, Phys. Rev. {\bf D 39} 2192 (1989).
\bibitem{mm} P. Mazur, E. Mottola, Phys. Rev {\bf D 64} 104022, 2001, Nucl. Phys. {\bf B 341}:187, (1990). E. Mottola, J. Math. Phys. {\bf 36} 2342 (1987).
\bibitem{adgloll} A. Dasgupta, R. Loll, Nucl. Phys. {\bf B606} 357, (2001).
\bibitem{bbm} Z. Bern, M. Blau, E. Mottola, Phys. Rev. {\bf D43}: 1212 , (1991).
\bibitem{adg2} A. Dasgupta, {\it The Euclidean path integral in quantum gravity} to appear in edited volume `Classical and Quantum Gravity Research' published by Nova Science Publishers. 
\bibitem{dewit}J. B. S. Dewitt, {\it Dynamical theory of groups and fields}, Gordon and Breach Science Publishers, (1965).
\bibitem{avramidi}I. G. Avramidi, {\it Heat Kernel and Quantum Gravity}, Lecture Notes in Physics, Springer, (2000). 
\bibitem{full} S. A. Fulling, {\it Aspects of Quantum Field Theory in Curved Space-Time}, Cambridge University Press, (1989).
\bibitem{bar} A. O. Barvinsky, G. A. Vilkovisky, Phys. Rept. {\bf 119} 1, (1985).
\bibitem{loll} R. Loll, {\it The emergence of spacetime or quantum gravity on your desktop}, Class. Quant. Grav. {\bf 25}: 114006 (2008)
\end{thebibliography}
\end{document}